\documentclass[aps,pre,showpacs,amsmath,amssymb,twocolumn,10pt,nofootinbib]{revtex4-1}
\usepackage{amsthm}
\usepackage{dsfont}
\pdfoutput=1
\usepackage{graphicx}
\newcommand{\psib}{\overline{\psi}}
\newcommand{\phib}{{\overline{\phi}}}
\newcommand{\chib}{{\overline{\chi}}}
\newcommand{\Psib}{{\overline{\Psi}}}
\newcommand{\beq}{\begin{equation}}
\newcommand{\eeq}{\end{equation}}
\usepackage{amsmath}
\usepackage{amsfonts}
\usepackage{epsfig}
\usepackage{slashed}
\usepackage{epstopdf}

\begin{document}

\title{Four fermion interactions in non abelian gauge theory }

\author{Simon Catterall, Aarti Veernala}
\affiliation{Department of Physics, Syracuse University, Syracuse, NY 13244, USA}

\date{\today}

\begin{abstract}
We continue our earlier study of the phase structure of a $SU(2)$ gauge theory whose action contains additional chirally invariant
four fermion interactions.
Our lattice theory uses a reduced staggered fermion formalism to generate two Dirac flavors in the continuum limit. In the current study we have tried
to reduce lattice spacing and taste breaking effects by using an improved fermion action incorporating stout smeared links. 
As in our earlier study we observe two regimes; for weak gauge coupling
the chiral condensate behaves as an order parameter differentiating a phase at small four fermi coupling where the condensate vanishes from a phase at strong
four fermi coupling in which
chiral symmetry is spontaneously broken. 
This picture changes qualitatively when the gauge coupling is strong enough to cause confinement; in this case we observe a first order phase transition
for some critical value of the four fermi coupling associated with a strong enhancement of the chiral condensate. We observe that
this critical four fermi coupling varies monotonically with bare gauge coupling - decreasing,
as expected, as the gauge
coupling is increased. We have checked that these results remain stable under differing levels of smearing. They appear to rule out the appearance of
new fixed points associated with chirally invariant four fermion interactions in confining non abelian gauge theories.
\end{abstract}


\maketitle
%------------------------------------------------

\section{Introduction}

The primary purpose of the Large Hadron Collider (LHC) is to probe the nature of electroweak symmetry breaking in the Standard Model (SM). In the SM this
is accomplished using a single scalar field - the Higgs. Unfortunately scalar field
theories are known to suffer from fine tuning and triviality problems and in consequence other extensions of the SM based on
supersymmetry or new strong dynamics have been put forth to provide more natural explanations of electroweak symmetry breaking. One of the simplest
realizations of dynamical electroweak symmetry breaking are technicolor models in which the SM is augmented with new fermions (techniquarks)  which interact
via  a new strong technicolor gauge force.  Strong interaction effects can then drive the formation of a techniquark chiral condensate which will in turn
precipitate electroweak symmetry breaking if the electroweak gauge group appears as a subgroup of the chiral group of the techniquarks
 \cite{TC-intro, TC-intro2}. While theories with QCD-like dynamics are experimentally excluded 
they are far less stringent constraints on these models if the dynamics deviates dramatically from QCD. This
is particularly true in light of the recent discovery of a light, Standard Model-like Higgs at the LHC.
One way to produce such a light state in a strongly coupled BSM theory is to consider models
lying close to the lower boundary of the conformal window. In this case it is conjectured
that one might be able to
identify the experimentally observed Higgs with a pseudo Goldstone boson - the dilaton - arising from
spontaneous breaking of approximate conformal symmetry.
Such theories are often termed walking gauge theories since the gauge
coupling should evolve slowly with energy scale \cite{TC-intro, walking-francesco1} and possess
other potentially useful features such as reduced contributions to the S parameter
and flavor changing neutral currents.

The search for such walking gauge theories has led to numerous recent lattice studies - see the conference reviews \cite{Fleming:2008gy, Pallante:2009hu, DelDebbio:2010zz,Fodor:2010zz}.  However, the simplest technicolor models, while giving rise to masses for the gauge bosons, are incapable of generating masses
for SM  fermions. To achieve this
more elaborate schemes such 
as extended technicolor (ETC) \cite{ETC-1,ETC-2,schrock2,schrock3} and top-condensation \cite{Miransky:1988xi,Miransky:1989ds,Bardeen:1989ds,Marciano:1989mj}
have been invoked.  At low energies these latter theories share a common framework; the actions contain additional four fermion
interactions coupling both SM fermions and techniquarks. Indeed most theories of BSM physics can be expected to yield
effective actions at low energy which contain additional four fermion terms.

In light of this it is important to study the structure of
non abelian theories in the presence of additional chirally invariant four fermion interactions.
This was the motivation behind our earlier work \cite{OurPaper}, where we explored the phase structure of a NJL model gauged
under the $SU(2)$ group. This model has also been examined using truncated Schwinger-Dyson
calculations in \cite{DSB-yamawaki} where strong four fermi interactions were claimed to lead to
new universality classes of model with potentially interesting consequences for model building.
Our work can also be seen as a non-perturbative check on these truncated Schwinger-Dyson results.

In our previous work we observed that the chiral condensate of the strongly coupled gauge theory underwent a strong enhancement for large values of the
four fermi coupling. The transition appeared very abrupt and therefore likely first order but we were unable to find definitive evidence of this in the earlier
work. In addition we observed that the critical four fermi coupling did not vary monotonically with the gauge coupling which led us to suspect that the results
of those simulations might be plagued with large lattice spacing effects.  To counter this we have revisited the model
using an improved fermion action in which the usual
gauge links are replaced by so called {\it stout smeared} gauge links. The smeared gauge fields are obtained by a gauge invariant
averaging procedure carried out over a local region in space. The essential idea is that this averaging procedure will tend to remove large UV fluctuations from the
lattice gauge field while preserving the IR physics. Smeared actions been shown to be very effective in lattice QCD \cite{anna-1,anna-2,MILC}. The specific scheme we employ was 
developed in \cite{Stout} for the case of $SU(3)$.

In the following section, we briefly describe our lattice theory and its connection to the gauged NJL model.  
Next we describe the method of stout smearing in the context of dynamical fermion simulations of an $SU(2)$ lattice
gauge theory. Finally, we move on to our numerical results, where we discuss the phase structure resulting from varying both gauge and four fermi
couplings. We show that suppressing lattice spacing effects via smearing removes some of the technical issues associated
with our earlier work but does not change the final conclusion: that the non abelian theories do not develop new fixed points
associated with strong four fermi interactions. Instead, 
the lattice theory undergoes a first order phase transition for large four fermi coupling. This
transition is associated with a rapid enhancement of the chiral condensate and the generation of large fermion masses. 

%------------------------------------------------
\section{Four fermion interactions - on and off the lattice}
\label{sec:themodel}

We will consider a model which consists of two flavors of massless Dirac fermions transforming in the
fundamental representation of a $SU(2)$ gauge group and coupled through an $SU(2)_L\times SU(2)_R$ chirally invariant four fermi interaction
of NJL type. To avoid potential sign problems we have utilized two copies of this basic 2 flavor doublet yielding a 
four flavor theory.
The action for a single doublet takes the form
\begin{eqnarray}
S &=& \int d^4x\; \psib ( i \slashed{\partial} -  \slashed{A}) \psi - \frac{G^2}{2N_f} [ (\bar{\psi} \psi)^{2} + (\bar{\psi} i \gamma_{5} \tau^{a} \psi )^{2} ]  \nonumber \\
&-& \frac{1}{2g^2}{\rm  Tr} [F_{\mu \nu} F^{\mu \nu}] ,
\label{eq:etcnjlaction}
\end{eqnarray}
where G is the four-fermi coupling, $g$ the usual gauge coupling
and $\tau^{a},a=1\ldots 3$ are the generators of the $SU(2)$ flavor group.

To implement the theory in Eq.~\ref{eq:etcnjlaction} on the lattice, it is convenient to re-parametrize the four fermi term in the continuum action via the use of scalar auxiliary fields. Specifically the fermion interaction term is replaced by the Yukawa terms
\begin{equation}
S_\text{aux} =\int d^4x\; \frac{G}{\sqrt{N_f}}\left(\bar{\psi}\psi\phi_4+\bar{\psi}i\gamma_5\tau^a\psi\phi_a\right)+\frac{1}{2}\left(\phi_4^2+\phi_a^2\right).
\end{equation}

We discretize the above action using a (reduced) staggered fermion formalism. This approach allows build a lattice theory which admits just
two continuum flavors of  fermion in the continuum limit without encountering rooting issues. Furthermore it allows us to write down
lattice four fermi terms which are invariant under a discrete subgroup of the continuum chiral symmetries~\cite{redstag-Smit-1,redstag-Smit-2}. The presence of four fermi interactions has an additional attractive feature - it allows us to study the lattice theory with exactly zero fermion mass where
these discrete symmetries are exact~\cite{Kogut:1996mj}. 

In order to do this, we first rewrite the fermionic sector of the continuum theory in terms of a new set of matrix valued fields, $\Psi$ and $\Psib$ which
naturally arise after twisting the original Lorentz symmetry with the flavor symmetry. We then expand these matrices on a basis corresponding to
products of gamma matrices and associate these products with staggered fields $\chi$, $\chib$.
\begin{eqnarray}
\Psi(x)&=&\frac{1}{8}\sum_b \gamma^{x+b}\chi(x+b) ,\\
\Psib(x)&=&\frac{1}{8}\sum_b (\gamma^{x+b})^\dagger\chib(x+b) ,
\label{matrixfields}
\end{eqnarray}
where $\gamma^{x+b}=\prod_{i=1}^4 \gamma_i^{x_i+b_i}$ and the sums correspond to the vertices in an elementary hypercube associated with lattice site $x$ as the components vary $b_i=0,1$. The lattice theory thus far  describes four flavors of continuum fermion.
To reduce the flavor content further it is possible for massless fermions 
to restrict  the single component fields $\chi$ and $\chib$ to even and odd sites, respectively. In this way we arrive
at the {\it reduced} staggered fermion action:
\beq
S= \sum_{x,\mu} \ \chi^{T}(x) \ \mathcal{U}_{\mu}(x) \ \chi(x + a_{\mu}) \ [\eta_{\mu}(x) +G\; \phib_\mu(x) \,\epsilon(x) \, \xi_\mu(x)],
\label{finalS-latt} \eeq where, 
\beq \mathcal{U}_{\mu} (x) = \frac{1}{2} [1+ \epsilon(x)] \; U_{\mu}(x) + \frac{1}{2} [1- \epsilon(x)] \; U_{\mu}^{*}(x) \label{mathcalU},\eeq 
\beq
\phib_\mu(x)=\frac{1}{16}\sum_b\phi_\mu(x-b)  \eeq and 
$\eta_\mu(x)$ is the usual staggered quark phase given by
\beq  \eta_{\mu}(x) = (-1)^{\sum_{1}^{\mu - 1} x_{\mu}} . \eeq
Notice that the action involves a single staggered field $\chi$ defined over all lattice sites but coupled through a modified gauge field ${\cal U}_\mu(x)$.
The two staggered tastes become the two physical quark flavors in the
continuum limit.  This action is invariant under the discrete shift transformations:

\begin{eqnarray}
\chi(x)&\to&\xi_\rho(x)\chi(x+\rho) , \\
U_\mu(x)&\to&U_\mu^{*}(x+\rho) , \\
\phi_\mu(x)&\to&(-1)^{\delta_{\mu\rho}}\phi_\mu(x+\rho) .
\end{eqnarray}
These shift symmetries
correspond to a {\it discrete} subgroup of
the continuum axial flavor transformations which act on the matrix field $\Psi$ according to
\beq \Psi\to \Psi\gamma_\rho\eeq

%%%%%%%%%%%%%%%%%%%%%%%%%%%%%%%%%%%%%%%%%%%%%%%
%%%%%%%%%%%%%%%%%%%%%%%%%%%%%%%%%%%%%%%%%%%%%%% 
%%%%%%%%%%%%%%%%%%%%%%%%%%%%%%%%%%%%%%%%%%%%%%%

\section{Stout Smearing}
\label{sec:smearing}

To reduce the magnitude of lattice spacing effects we replace the links $U_\mu(x)$ used in constructing the staggered fermion action by stout smeared
links $U_\mu^\prime(x)$ which are defined as
\begin{equation}
U_{\mu}^{\prime}(x) =  \{ e^{Q_{\mu}(x)} \} \, U_{\mu}(x).
\label{eq:linksmearing}
\end{equation} where the traceless, antihermitian matrix $Q_{\mu}$ is given by
\beq
 Q_{\mu}(x) = \frac{1}{2} \, [ \Omega_{\mu}^{\dagger}(x) - \Omega_{\mu}(x) ] - \frac{1}{2N} \, {\rm Tr} [ \Omega_{\mu}^{\dagger}(x) - \Omega_{\mu}(x) ]. 
 \label{eq:qmu}
 \eeq
 with
 \beq
\Omega_{\mu}(x) = C_{\mu}(x) \, U^{\dagger}_{\mu}(x),
 \label{eq:omegamu}
 \eeq 
 and
 \begin{widetext} \beq
 C_{\mu} = \sum_{\mu \neq \nu} \: \rho \; [  U_{\nu}(x) \, U_{\mu}(x + \nu) \, U_{\nu}^{\dagger}(x + \mu) + U_{\nu}^{\dagger}(x - \nu) \, U_{\mu}(x - \nu) \, U_{\nu}(x - \nu + \mu) ],
\label{eq:Cmu}
\eeq \end{widetext} The parameter $\rho$ can be tuned to optimize the smearing - in this work we have set $\rho=0.1$. 
To leading order in $\rho$ it is easily seen that this procedure generates a smeared link which
is a weighted sum over the original link and its surrounding staples.
In practice this procedure is iterated with the level n smeared fields $U_\mu^{(n)}$ being constructed from the fields $U_\mu^{(n-1)}$ at level n-1 via the obvious
relation
\begin{equation}
U_\mu^{(n)}(x)=\{ e^{Q_\mu^{(n-1)}(x)}\}\, U_\mu^{(n-1)}(x)
\label{finalsmearedlink}
\end{equation}
 
We have run our simulations using $n=1$ and $n=2$ levels of smearing.
For details on how this smearing in implemented  and the calculation of the smeared force term required by the RHMC simulation
we refer the reader to the Appendix.

\section{Numerical results}
Upon integration over
the basic fermion doublet we obtain a Pfaffian ${\rm Pf(M(U))}$ depending in a complicated way on the
original gauge field $U_\mu(x)$\footnote{Note that the fermion operator appearing in eqn.~\ref{finalS-latt} is antisymmetric and hence a Pfaffian rather
than a determinant results after integrating over the single fermion field $\chi$}.
The required pseudofermion weight for 2 such doublets ($N_f=4$  flavors) is then
${\rm Pf}(M)^2$. The pseudoreal character of
$SU(2)$ allows us to show that this
Pfaffian is purely real  and so the four flavor simulations suffer from no sign problem. In practice we implement this weight in the path
integral using a
pseudofermion operator of the form  $(M^\dagger M)^{-{\frac{1}{2}}}$ and a RHMC algorithm.
A standard Wilson gauge action is employed for the gauge fields. 

We have focused our study on 
a lattice of size $6^4$ and a bare fermion mass $m=0$.  We have simulated a range of gauge couplings $\beta \equiv 4/g^2=1.8-2.2$ with additional
runs at $\beta=10.0$ to test the pure NJL model in which the gauge interactions are switched off. For each gauge coupling we have performed an extensive
scan over a dozen or so values of the four fermi coupling $G$.
\begin{figure}[htb]
\begin{center}
\includegraphics[height=62mm]{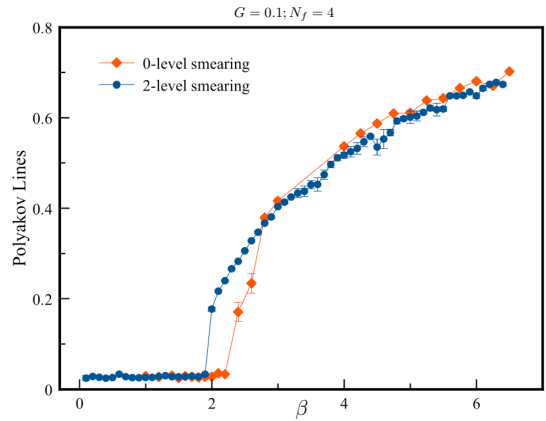}
\caption{Polyakov line vs $\beta$ at $G=0.1$ for four flavours}
\label{poly-L468}
\end{center}
\end{figure}
The range of $\beta$ that was employed
in the simulations was determined by examining the
average plaquette and Polyakov line as $\beta$ was varied holding
the four fermi coupling fixed at $G=0.1$.
In figure ~\ref{poly-L468}.
we see a strong finite volume phase transition for $\beta>2.2$ on a
lattice with $L=6$ and no smearing.  Furthermore, this transition occurs at smaller bare gauge
coupling, $\beta>1.9$, when  two levels of stout smearing are employed. 

The primary observable used in this
study is the chiral condensate which
is computed from the gauge invariant one link operator
\begin{widetext}\beq
\sum_x \left\langle \chi (x)\left({\mathcal U}_\mu(x)\chi (x+e_\mu)+{\mathcal U}^\dagger_\mu(x-e_\mu)\chi(x-e_\mu)\right) \right\rangle \epsilon (x)\xi_\mu(x)\eeq \end{widetext}
Because of the absence of spontaneous
symmetry breaking in finite volume we measure the absolute value of this operator. In a chirally broken phase we
expect this to approach a constant as the lattice volume is sent to infinity. Conversely if chiral symmetry is restored
this observable will approach zero in the same limit. We have also used  the auxiliary scalar field to monitor this observable.

\begin{figure}
\begin{center}
\includegraphics[height=70mm]{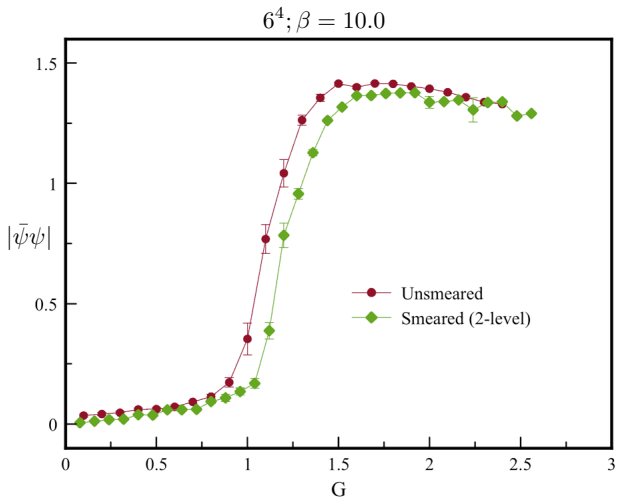}
\caption{$| \langle\chib\chi\rangle |$ vs G for $\beta=10.0$,  $6^4$ lattice}
\label{psibpsi-L6N4-beta10-alllevel}
\end{center}
\end{figure}

Having used these simple gauge observables to determine an appropriate range for the gauge coupling $\beta$ we now turn to the behavior of the
system as the four fermi coupling is varied.
As a benchmark we have examined the system for a small value of the gauge coupling at $\beta=10.0$ \footnote{Note: gauge coupling, $g^{2}$ varies as $\frac{1}{\beta}$.}. The condensate is
shown in figure~\ref{psibpsi-L6N4-beta10-alllevel} as the four fermi coupling varies. The plot reveals evidence for a phase
transition separating a chirally symmetric phase at small $G$ from a broken phase for $G>G_c$. The relatively smooth
behavior is consistent with earlier work using sixteen flavors of naive fermion reported
in \cite{annakuti} which identified a line of second order phase transitions in this region of
parameter space. It also agrees with the behavior seen in previous simulations of staggered quarks \cite{Hands:1997uf}.

\begin{figure}
\begin{center}
\includegraphics[height=70mm]{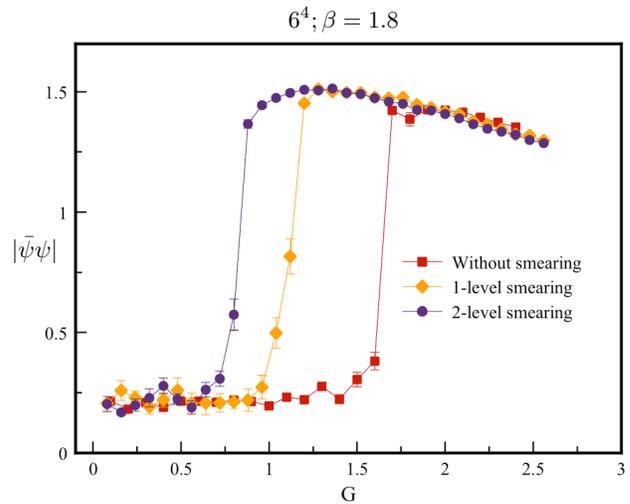}
\caption{$\langle\chib\chi\rangle$ vs $G$ at $\beta=1.8$ for $6^4$ lattice}
\label{psibpsi-withnosmear}
\end{center}
\end{figure}

We now turn to our results at non-zero gauge coupling, corresponding to smaller values of $\beta$ (stronger gauge coupling).
Figure \ref{psibpsi-withnosmear} shows results for the condensate 
in the confining theory at $\beta=1.8$ for several levels of smearing.
Notice that the chiral condensate is now non-zero even for vanishing four fermi coupling
consistent with spontaneous chiral symmetry breaking in the pure gauge
theory. 
However, it jumps abruptly to much
larger values when the four fermi coupling exceeds some critical value. Notice that while the limiting values of the condensate in both
small $G$ and large $G$ phases do not depend on how much smearing is done, the position of the transition itself depends on the smearing level. This should not
be particularly surprising; the magnitude of critical couplings associated with lattice phase transitions generically depend on details at the scale of the lattice spacing. 
However, notice that the
chiral condensate, being an IR quantity, does not show this sensitivity.

\begin{figure}
\begin{center}
\includegraphics[height=75mm]{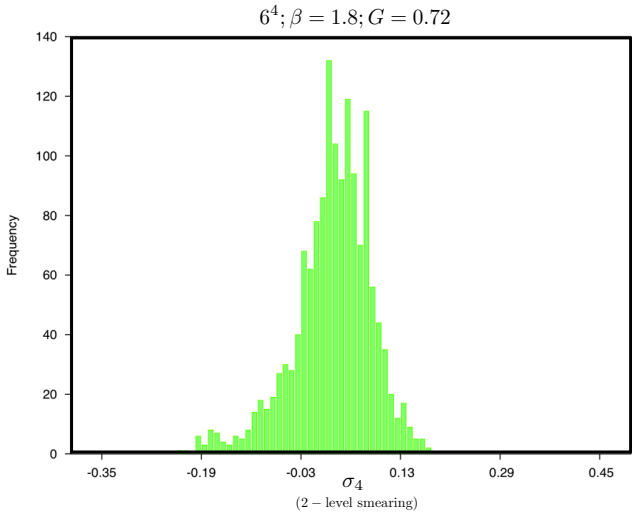}
\caption{Histogram of $\sigma_4$ for G=0.72.}
\label{hist-G0pt72}
\end{center}
\end{figure}

\begin{figure}
\begin{center}
\includegraphics[height=75mm]{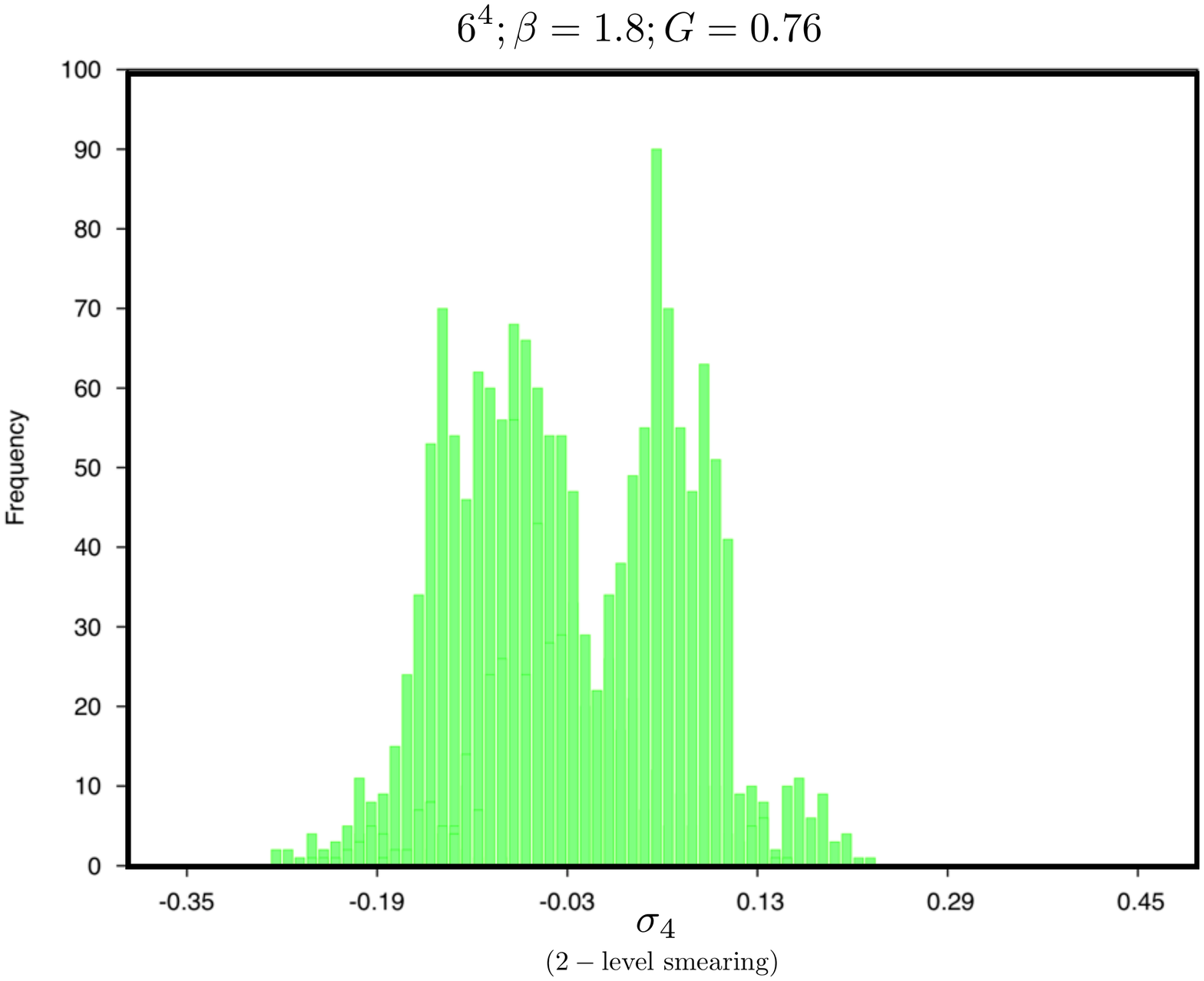}
\caption{Histogram of $\sigma_4$ for G=0.76.}
\label{hist-G0pt76}
\end{center}
\end{figure}

\begin{figure}
\begin{center}
\includegraphics[height=75mm]{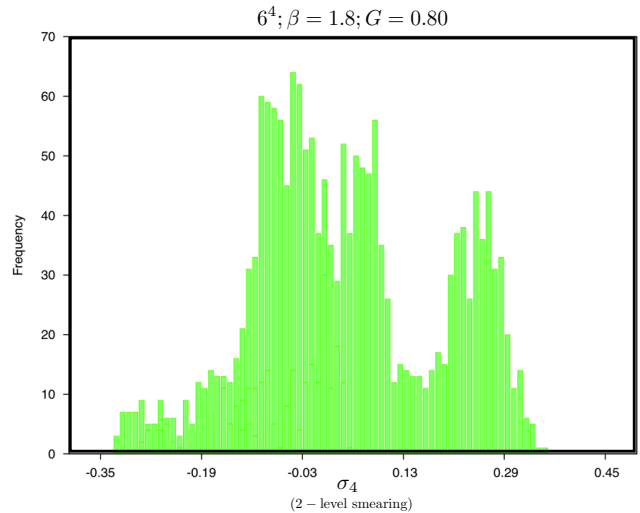}
\caption{Histogram of $\sigma_4$ for G=0.80.}
\label{hist-G0pt80}
\end{center}
\end{figure}

\begin{figure}
\begin{center}
\includegraphics[height=75mm]{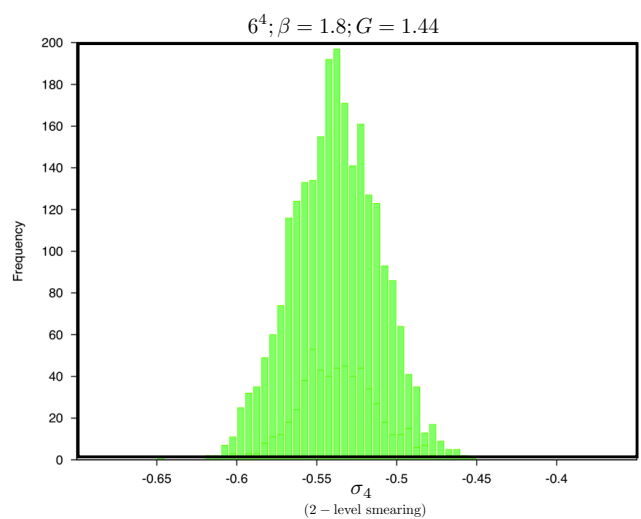}
\caption{Histogram of $\sigma_4$ for G=1.44.}
\label{hist-G1pt44}
\end{center}
\end{figure}

Figure~\ref{psibpsi-withnosmear} appears to show that the jump in the chiral condensate
is markedly discontinuous in character - reminiscent of a first order phase transition. 
Further evidence for this first-order behavior is seen in figures ~\ref{hist-G0pt72},\ref{hist-G0pt76}, \ref{hist-G0pt80}, and \ref{hist-G1pt44} which show histograms of the
condensate (as measured by the scalar field average $\sigma_4$) for several four fermi couplings
$G$ around the phase transition.
The histograms tell a clear story; for $G$ both below and above the transition a single peak is seen.
In contrast for values of $G$ close to the jump in the condensate the double
peak structure indicates that condensate has tunneled back and forth between two
different  states over the course of the run. Such a tunneling is clear indication of the presence of a first-order transition in the lattice system.

\begin{figure}
\begin{center}
\includegraphics[height=70mm]{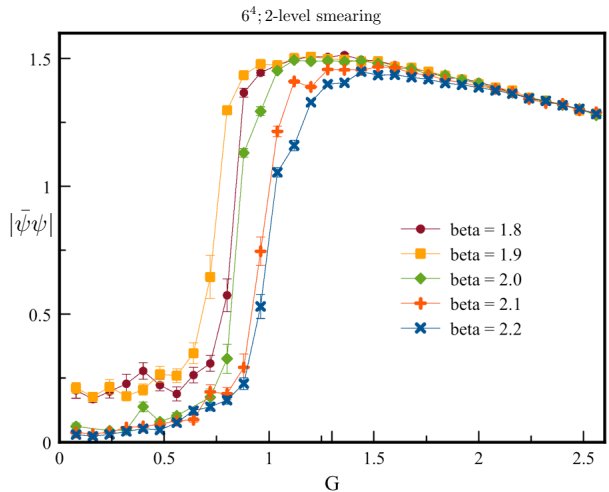}
\caption{$\langle\chib\chi\rangle$ vs $G$ for varying $\beta$. 2-level smearing has been implemented here.}
\label{psibpsi-allbeta.pdf}
\end{center}
\end{figure}
We have also examined the dependence of these features on the magnitude of the gauge coupling $\beta$.
Figure~\ref{psibpsi-allbeta.pdf}, shows a plot of the condensate 
vs $G$ for $N_{f} = 4$ for a fixed lattice size $L=6$ and using two levels of smearing.  Two regimes are clearly visible; for $\beta<1.9$ or so the chiral
condensate is non zero for small $G$ corresponding to a strongly coupled, confining and chirally broken gauge theory even in the absence of four fermi terms. Conversely
for $\beta>2.0$ the chiral condensate goes to zero for vanishing four fermi coupling and the system is deconfined in the absence of the four fermi 
coupling. This is not a surprise; it corresponds to the usual deconfinement one sees when the lattice size becomes smaller than
the confinement scale.

Notice that the onset of the crossover occurs for smaller values of the critical four-fermi coupling, $G_{cr}$, as we increase the strength of the gauge coupling (corresponding to decreasing $\beta$). This monotonic dependence of the critical four fermi coupling  contrasts with the situation
seen earlier in our study with an unsmeared action \cite{OurPaper}.  In the strong coupling regime, since the chiral symmetry is already broken, the cross over occurs quickly at a small (weak) four-fermi coupling. As one approaches the weak coupling limit, the model is less influenced by the gauge interactions and the chiral symmetry remains intact until the four-fermi coupling becomes strong enough to break the symmetry. 

\section{Summary}

In this paper we have conducted numerical simulations
of a four flavor non abelian gauge theory in the presence of additional chirally invariant
strong four fermion interactions. The latter are expected to be present in generic effective theories
of BSM physics including (walking) technicolor, top quark condensation and little Higgs theories. 
We employ a reduced staggered fermion lattice action which preserves a discrete subgroup of the continuum chiral
symmetry. In addition we have improved this action  using stout smearing techniques.

We have examined the model for a modest lattice size of $6^4$ and over a range of gauge couplings $\beta$, and four fermi interaction strength $G$. 
In the NJL limit $\beta\to\infty$
we find evidence for a continuous phase transition corresponding to
the expected spontaneous breaking of chiral symmetry due to strong four fermi interactions. The location and character of this
transition is, not surprisingly, insensitive to the smearing. However, for gauge couplings $\beta$ that
generate confinement and a non-zero chiral condensate even at $G=0$, this transition  appears
much sharper and indeed we provide strong evidence that the transition is first order. 
Such a phase transition precludes the existence of any new continuum limits in the
model.
The location of this transition occurs at a critical four-fermi coupling which depends on the
level of smearing and decreases monotonically with increasing gauge coupling.

Thus our results are consistent with the idea that the second order phase transition which exists in the pure
NJL theory ($G=\infty$) survives in the presence of weak gauge interactions. However our results indicate that this line of transitions becomes
first order at a point at which the gauge coupling becomes strong enough to cause confinement. In a finite volume this occurs at a finite $\beta$ but
presumably this first order region extends all the way to $\beta=\frac{1}{g^2}=\infty$ in infinite volumes.
The gauged NJL model we study in this paper was examined using
a truncated Schwinger-Dyson analysis in ~\cite{DSB-yamawaki} in which it was claimed that the phase diagram of the gauged NJL
model should include a new critical line along which critical exponents such as the mass anomalous dimension vary continuously. 
Our results 
are {\it inconsistent} with this scenario but are consistent 
with the approximate analysis performed in ~\cite{Takeuchi:1989qa} which argues for a strong enhancement of
the chiral condensate for large four fermi coupling. While our results seem to exclude new universality classes 
due to strong four fermi couplings in confining gauge theories it is still logically possible that this
conclusion might be very different in a gauge theory which is conformal at $G=0$.

\acknowledgments This work is supported in part by DOE grant
DE-FG02-85ER40237. Simulations were carried on USQCD facilities at Fermilab.

\appendix
\section{Calculation of the stout smeared force in $SU(2)$}
Stout smearing is implemented in our code in two parts. The first involves replacing the gauge links in the fermionic action by the smeared links, ${U_{\mu}^{(n)}}(x)$ as given in equation \ref{finalsmearedlink}. The second, and the non-trivial, part of implementing smearing lies in the calculation of the smeared force $F_{\mu}(x)$ needed in the auxiliary molecular dynamics evolution which forms the kernel of
the dynamical fermion algorithm.
The force $F_{\mu}(x)$ is defined as:
\beq
F_{\mu}(x)=U_{\mu}(x) \frac{\partial S(U^{(n)})}{\partial U_{\mu}(x)}
\label{forcedef}
\eeq 
The complication that arises at this point is that we have no explicit representation of the smeared action in terms of the original gauge links - indeed such an expression
would be very complicated given that the smeared links are determined iteratively. Instead we follows the approach of \cite{Stout} and derive an iterative
expression for the smeared force by considering the rate of change of the action over a fictitious (simulation) time $t$:
\beq
\frac{\partial}{\partial t}S(U^{(k)}) = 2\;Re \sum_{\mu,x} \; {\rm Tr} \left[ f^{(k)}_{\mu}(x) \frac{\partial}{\partial t}U^{(k)}_{\mu}(x) \right].
\label{Srate}
\eeq Here, $f^{(k)}_{\mu}(x)=\frac{\partial S(U^{(k)})}{\partial U^{(k)}_{\mu}(x)}$. In order to simplify notation, we denote the $k$ level fields with a prime and the $(k-1)$ level fields without a prime. Using  equation (\ref{eq:linksmearing}), we have,
\beq
\frac{\partial }{\partial t}U'_{\mu}(x)=\left(\frac{\partial}{\partial t}e^{Q_{\mu}(x)}\right)U_{\mu}(x) + e^{Q_{\mu}(x)}\left(\frac{\partial U_{\mu}(x)}{\partial t}\right).
\label{smUrate}
\eeq 
The matrix exponentials can be handled using the $SU(2)$ identities:
\beq
e^{Q}=(\cos q)I_{2X2} + \frac{\sin q}{q}Q,
\label{su2def}
\eeq where q the eigenvalues of the traceless antihermitian matrix $Q$ are given by
\beq
q^{2} = -\frac{1}{2}{\rm Tr}(Q^{2})
\label{q}
\eeq
Using this the derivatives of $e^Q$ can be expressed as
\beq
\frac{\partial}{\partial t}e^{Q_{\mu}(x)}= {\rm Tr} \left[ Q_{\mu}(x)\frac{\partial Q_{\mu}(x)}{\partial t} \right] B + \frac{\sin{q}}{q}\left(\frac{\partial Q_{\mu}(x)}{\partial t} \right).
\label{exprate2}
\eeq 
with
\beq B_\mu=\frac{\sin{q}}{2q} -\frac{(\cos{q}-\sin{q}/q)}{2q^{2}}Q_{\mu}, \eeq 
Using these results eqn.~\ref{Srate} can be expressed as
\begin{widetext}\beq
\sum_{\mu,x} \; {\rm Tr} \left[ f'_{\mu}(x) \frac{\partial}{\partial t}U'_{\mu}(x) \right] = \sum_{\mu,x} \left[{\rm Tr} \left( f'_{\mu}(x) e^{Q_{\mu}(x)}\frac{\partial}{\partial t}U_{\mu}(x)\right) -{\rm Tr} \left(\Lambda_{\mu}(x)\frac{\Omega_{\mu}(x)}{\partial t}\right) \right].
\label{Srate2}
\eeq \end{widetext} where

\beq
\Lambda_{\mu}(x) = - \left[\Gamma^{\dagger}_{\mu}(x) - \Gamma_{\mu}(x) \right] + \frac{1}{N} {\rm Tr} \left[\Gamma^{\dagger}_{\mu}(x) - \Gamma_{\mu}(x) \right],
\label{lambda}
\eeq
and
\beq
\Gamma_{\mu}(x) = {\rm Tr} \left[f'_{\mu}(x)B_{\mu}(x)U_{\mu}(x) \right] Q_{\mu}(x) + \frac{\sin{q}}{q}U_{\mu}(x)f'_{\mu}(x)
\label{gamma}
\eeq and $\Omega_{\mu}(x)$ is given by equation ($\ref{eq:omegamu}$). Using the cyclic properties of the trace we ultimately obtain,

\begin{eqnarray}
f_{\mu}(x)&=& f'_{\mu}(x)(\cos{q}I_{2X2}+\frac{\sin{q}}{q}Q) - C^{\dagger}_{\mu}(x)\Gamma_{\mu}(x)\nonumber \\
                 &+& \rho \sum_{\mu,x} \Theta_{\mu\nu}(x)\label{fmu}\end{eqnarray}
%\label{fmu}\end{eqnarray}
where,

\begin{eqnarray}
\Theta_{\mu\nu}(x)=U_{\nu}(x+\mu)U^{\dagger}_{\mu}(x+\nu)U^{\dagger}_{\nu}(x)\Lambda_{\nu}(x) \nonumber \\
+ U^{\dagger}_{\nu}(x+\mu-\nu)U^{\dagger}_{\mu}(x-\nu)\Lambda_{\nu}(x-\nu)U_{\mu}(x-\nu) \nonumber \\
- U^{\dagger}_{\nu}(x-\mu+\nu)U^{\dagger}_{\mu}(x-\nu)\Lambda_{\nu}(x-\nu)U_{\nu}(x-\nu) \nonumber \\
- \Lambda_{\nu}(x+\mu)U_{\nu}(x+\mu)U^{\dagger}_{\mu}(x+\nu)U^{\dagger}_{\nu}(x) \nonumber \\
+ U_{\nu}(x+\mu)U^{\dagger}_{\mu}(x+\nu)\Lambda_{\mu}(x+\nu)U^{\dagger}_{\nu}(x) \nonumber \\
+ U^{\dagger}_{\nu}(x+\mu-\nu)\Lambda_{\nu}(x+\mu-\nu)U^{\dagger}_{\mu}(x-\nu)U_{\nu}(x-\nu).  \nonumber \\
\end{eqnarray} 
Eqn. \ref{fmu} gives $f_\mu(x)\equiv f_\mu^{(k-1)}(x)$  in terms of $f^\prime_\mu(x)=f^{(k)}_\mu(x)$ at level $k$. 
Given an initial ``naive'' force at level $n$ $f^{(n)}_\mu=\frac{\partial S(U^{(n)}}{\partial U_\mu^{(n)}}$ it may be iterated to yield $f^{(0)}_\mu(x)$ and hence the
final smeared force needed by the molecular dynamics equations
\beq
F_{\mu}(x)=U_{\mu}(x) f^{(0)}_{\mu}(x),
\label{Fdef}
\eeq 

%\begin{acknowledgments}
%SMC is supported in part by DOE grant
%DE-FG02-85ER40237. The simulations were carried out using USQCD
%resources at Fermilab and Jlab.
%\end{acknowledgments}

%\bibliographystyle{ieeetr}
%\bibliography{gaugedNJL}

%\bibliography{4fermiPRDNotes}

\end{document}